\documentstyle[aps,prl,floats,epsf]{revtex}
\epsfclipon
\newcommand{\smeq}{\! \! = \!}

\newcommand{\cl}{\chi_{\scriptscriptstyle L}}
\newcommand{\Ef}{E_{\scriptscriptstyle F}}
\newcommand{\kf}{k_{\scriptscriptstyle F}}
\newcommand{\vf}{v_{\scriptscriptstyle F}}
\newcommand{\pf}{p_{\scriptscriptstyle F}}
\newcommand{\lambdaf}{\lambda_{\scriptscriptstyle F}}
\newcommand{\br}{{\bf r}}

\newcommand{\G}{{\cal G}}   

\draft

\begin{document}

\twocolumn[{   

\title{\bf
Chaos and Interacting Electrons in Ballistic Quantum Dots
}

\author{Denis~Ullmo,$^{1,2}$ Harold~U.~Baranger,$^{1}$ Klaus~Richter,$^{3}$
Felix~von~Oppen,$^{4}$ and Rodolfo~A.~Jalabert$^{5}$}

\address{$^{1}$Bell Laboratories--Lucent Technologies, 700 Mountain Avenue,
Murray Hill, New Jersey 07974-0636}

\address{$^{2}$Division de Physique Th\'eorique, 
Institut de Physique Nucl\'eaire, 91406 Orsay Cedex, France}

\address{$^{3}$Max--Planck--Institut f\"ur Physik
komplexer Systeme, N\"othnitzer Str.\ 38, 01187 Dresden, Germany}

\address{$^{4}$Department of Condensed Matter Physics, Weizmann Institute of
Science, 76100 Rehovot, Israel}

\address{$^{5}$Universit\'e Louis Pasteur, IPCMS-GEMME,
23 rue du Loess, 67037 Strasbourg Cedex, France}

\date{August 12, 1997}

\maketitle
\mediumtext
{\tighten
\begin{abstract}
We show that the classical dynamics of independent particles can determine
the quantum properties of interacting electrons in the ballistic regime. 
This connection is established using diagrammatic perturbation theory and 
semiclassical finite-temperature Green functions. 
Specifically, the orbital magnetism is greatly enhanced over the Landau 
susceptibility by the combined effects of interactions and finite size. 
The presence of families of periodic orbits in regular systems makes their 
susceptibility parametrically larger than that of chaotic systems, a 
difference which emerges from correlation terms.
\end{abstract}
}
\vspace{0.3truein}

}]  

\narrowtext


The connection between classical dynamics and wave interference has
attracted attention in many fields of physics recently \cite{LesHqchaos},
including atomic, mesoscopic, and optical physics. A central question is
to what extent the quantum properties of classically regular and chaotic
systems differ.  On the whole, this question has been addressed for
non-interacting systems. It is now known that many quantum properties are
in fact strongly influenced by the nature of the classical dynamics-- the
density of states, the quantum corrections to the conductance, and the
optical absorption to name a few.

We wish to address this question for {\it interacting} systems and, in
particular, to investigate the role of the classical dynamics of the
non-interacting system in this context.  If the interactions are strong,
the non-interacting classical dynamics will be of little relevance.
However, if the interactions are short-range and not too strong, the
non-interacting classical dynamics may be important, and its role can be
assessed with perturbation theory. This regime is physically relevant: it
applies to a high density two-dimensional electron gas in which the
quasi-particles interact weakly through the short-range screened Coulomb
interaction. We find that at {\it first} order in the interaction there is
a difference between regular and chaotic systems, but one which is only
numerical, not qualitative. Intriguingly, as the perturbation theory is
carried out to {\it higher} orders a qualitative difference emerges:
thermodynamic properties scale differently with Fermi energy for chaotic
and regular systems.  This correlation effect shows that the nature of the
classical dynamics can have a substantial effect on the quantum properties
of an interacting system.

To be specific, we study the magnetic response of an ensemble of ballistic
quantum dots formed from a two-dimensional electron gas. Recent
fabrication progress has made possible phase-coherent electronic
microstructures much smaller than the mean free path.  In these
``ballistic'' quantum dots, one can think of electrons moving along
straight lines between specular reflections off the confining potential.
Because this motion is qualitatively different from that taking place in
bulk materials, a variety of new behavior has been observed experimentally
\cite{RevMes}.  In particular, the magnetic susceptibility of an ensemble
of ballistic squares has been measured \cite{levy93}, and a large
enhancement over the Landau response $\cl$ was found.  First attempts to
understand this experiment within non-interacting models pointed to the
importance of the classical dynamics \cite{levy93,vop95,urj95}. The
inclusion of interactions in such systems is our main concern, though much
of the discussion applies to ballistic structures in general.

For the magnetic response, the high-density expansion (RPA) of the
thermodynamic potential \cite{AGD} has to be extended by including
Cooper-like correlations, as carried out previously for disordered metals
\cite{aslamazov,doubleA,ambegaokar}.  Such expansions are typically used
beyond the high-density limit and yield reliable results for the bulk
provided some sets of terms are properly resummed. We continue to follow
this point of view for quantum dots, where the ``small parameter'' $r_s
\smeq r_0 / a_0$ is about 2. ($\pi r_0^2$ is the average area per
electron, and $a_0$ is the Bohr radius in the material.) We show that
these expansions are particularly insightful when combined with a
semiclassical approximation from which the connection to the nature of the
classical dynamics can be made.  Thus, we will assume that $\kf a \gg 1$
($a$ is the size of the microstructures and $\kf$ the Fermi wavevector)
and that the magnetic field $B$ is classically weak (cyclotron radius $\gg
a$).


{\it Semiclassical approach.--} The perturbation expansion 
\cite{AGD,doubleA,ambegaokar} for the interaction contribution to the 
thermodynamic potential $\Omega$ yields the magnetic susceptibility
through $\chi \equiv (-1/a^2) \partial^2 \Omega / \partial B^2$. 
The dominant terms are shown in Fig. 1. The
screened Coulomb interaction (wavy lines) is treated as local
\cite{localint}, $U(\br - \br') = \lambda_0 N(0)^{-1} \delta (\br -
\br')$, with $N(0)$ the density of states and $\lambda_0 = 1$
identifying the order of perturbation.  Straight lines represent the
``free'' finite-temperature Green function in the presence of the
confining potential,
\begin{eqnarray}
   \G_{\br, \br'}(\epsilon_n) \!= \theta (\epsilon_n)G^R_{{\bf r},{\bf r'}}
   (\Ef \!+\!i\epsilon_n) + \theta(-\epsilon_n) G^A_{{\bf r}, {\bf r'}}
   (\Ef \!+\!i\epsilon_n) \;.\nonumber
\end{eqnarray}
Here, $\Ef$ is the Fermi energy,
$\epsilon_n \smeq (2n+1) \pi / \beta$ the Matsubara frequencies, 
and $G^{R,A}$ the retarded, advanced Green functions related by
$G^A_{\br,\br'}(E)=[G^R_{\br',\br}(E^*)]^*$.

\begin{figure}
\epsfxsize=8.5cm
\epsffile[  50 410 550 530 ]{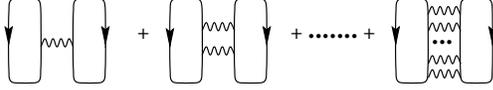}
\caption{
Leading Cooper-channel diagrams for the interaction contribution to the 
thermodynamic potential. 
}
\end{figure}

Semiclassically, $G^R$ is the sum of the contributions
$G^{R;j}_{{\bf r},{\bf r'}}$ of each classical trajectory $j$ 
from $\br$ to $\br'$ \cite{LesHqchaos}: in 2D
\begin{equation} 
   \label{GRA} G^R_{\br, \br'}(E) \simeq 
   {1 \over \sqrt{ 2\pi(i\hbar)^3}} 
   \sum_{j : {\bf r} \to {\bf r'}} 
   D_j \, e^{iS_j/\hbar - i\pi\nu_j/2}  
\end{equation} 
where $S_j \smeq \int_{\bf r}^{\bf r'} {\bf p} \cdot d{\bf r}$ is the
classical action of trajectory $j$, $D_j^2 \smeq (\dot x \dot x')^{-1}
\left| {\partial^2 S_j / \partial y \partial y'} \right|$ the
classical density, and $\nu_j$ a Maslov index.  Using $(\partial S_j /
\partial E) \smeq t_j$ and $(\partial S_j /\partial B) \smeq (e  / c) A_j $,
where $t_j$ and $A_j$ are the traversal time and area, 
one finds that the contribution of trajectory $j$ to $G^R$ is
\begin{eqnarray}
   G^{R;j}_{{\bf r},{\bf r'}}(\Ef\!+&&\!i\epsilon_n,B)=G^{R;j}_{{\bf r},
   {\bf r'}}(\Ef,B\!=\!0)\nonumber\\
   &&\times\exp\left[ -\epsilon_n t_j/\hbar\right] \times \exp
   \left[ i 2 \pi B A_j/\phi_0 \right]
\end{eqnarray}
where $\phi_0\!=\!h c / e $ is the flux quantum. 
Note that temperature introduces time and length scales 
$ t_T\!=\!L_T/\vf\!=\!\hbar\beta / \pi$ which exponentially suppress the 
contributions of long paths through the term
$\epsilon_n t_j / \hbar\!=\!(2n\!+\!1) t_j / t_T$.
($\vf$ is the Fermi velocity of a billiard.) This provides a complete 
description in the semiclassical perturbative regime.

We start with the first-order (Hartree-Fock) term in the diagrammatic
expansion
\begin{equation} 
   \label{first} 
   \Omega^{(1)} = { \lambda_0 \over \beta}
   \sum_\omega{\rm Tr}\, \{ \Sigma_{\br, \br'}(\omega) \}
\end{equation} 
where the trace refers to the spatial arguments of the
particle-particle propagator \cite{AGD}
\begin{equation} 
   \label{sigma} 
   \Sigma_{\br, \br'}(\omega)={1\over \beta
   N(0) } \sum_{\epsilon_n}^{\Ef} {\cal G}_{{\bf r}, {\bf r'}}
   (\epsilon_n) {\cal G}_{{\bf r}, {\bf r'}}(\omega-\epsilon_n) \; .  
\end{equation} 
($\omega \sim \omega_{\tilde n} = 2 \tilde n \pi / \beta$).  
The short-length (high-frequency) behavior is incorporated in the
screened interaction, thus requiring a cutoff of the frequency sums at
${\Ef}$ \cite{doubleA}. Semiclassically, 
$\Sigma_{{\bf r}, {\bf r'}}$ is a
sum over pairs of trajectories joining $\br$ to $\br'$.  However, most
pairs yield highly oscillating contributions 
which, after the spatial integrations, give higher order terms in $1/\kf a$.
To leading order, only those pairs contribute to the susceptibility
whose dynamical phases $\exp[i S_j(B\!=\!0)/\hbar]$ cancel while
retaining a magnetic-field dependence.  This is achieved by pairing
each orbit $j$ with its time reverse.  The trace in
Eq.~(\ref{first}) yields a sum over closed but not necessarily periodic
trajectories (see  Fig.~2, left, for a square).
This ``diagonal'' or ``Cooper channel'' is present independent of the 
nature of the classical dynamics \cite{disorder}, and we
will return to it below.  We first turn to an additional contribution 
present for integrable systems which is central to this paper. 

\begin{figure}
\epsfxsize=8.5cm
\epsffile[  50 410 580 610 ]{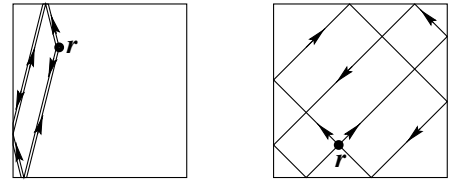}
\caption{ 
Typical pairs of real-space trajectories that contribute to the 
average susceptibility to first order in the interaction in the 
diagonal channel (left) and the non-diagonal channel (right).}
\end{figure}

{\it Non-diagonal channel.--} In integrable systems, periodic orbits
come in families within which the action integral is constant.  
If, as is generally the case, in some part of configuration space 
two orbits of the same family cross at a given point, 
it is possible to cancel the dynamical phases by pairing them (Fig.~2, right).  
This non-diagonal first-order contribution involves a term for each family
of periodic orbits. For the square billiard, and at not too low temperature
($L_T \lesssim 2a$), only the shortest of these
periodic orbits has to be taken into account, namely the family (11) with
length $L_{11} \smeq 2\sqrt{2}a$ shown in Fig.~2 (right). 
In this case, we find for the susceptibility
\begin{equation}
   \label{nondiag} 
   {\langle \chi^{\rm non-diag}_{11} \rangle \over \cl} =
   - \ {3 \kf a \over 2 (\sqrt{2}\pi)^3} \ 
   \ { d^2 {\cal C}^2(\varphi) \over d \varphi^2} 
   \ R^2\left(\frac{L_{11}}{L_T}\right) \; ;
\end{equation} 
the temperature dependence is governed by the function $R(x) \smeq {x /
\sinh(x)}$ and the field dependence by $ {\cal
C}(\varphi) \smeq ({2 \varphi})^{-1/2} \left[ \cos(\pi
\varphi) {\rm   C}(\sqrt{\pi   \varphi})  +  \sin(\pi  \varphi)   {\rm
S}(\sqrt{\pi \varphi}) \right]$, with $\varphi \smeq Ba^2  / \phi_0 $, and
${\rm C}$  and ${\rm S}$ Fresnel functions. As in the non-interacting
case \cite{vop95,urj95}, the contribution Eq.~(\ref{nondiag}) is
linear in $\kf  a$ and has a temperature scale related to the length of the 
periodic orbit. Quantitatively, the non-diagonal contribution
of the family (11) and its repetitions is shown as the dashed curve in 
Fig. 3. {\it Thus the existence of a family of periodic orbits-- a
characteristic of the non-interacting classical dynamics-- is associated
with an additional first-order interaction contribution to the
susceptibility.}

Higher-order terms in perturbation theory also contain
non-diagonal contributions. However, in these terms the location of
the additional interaction points is severely limited: they must lie
on both periodic orbits to cancel the dynamical phases and so must be 
near the intersections of the two orbits. Further analysis shows that
these contributions are therefore smaller by a factor of $1/\kf a$.  By
contrast, we will now show that the diagonal contribution is strongly
renormalized by higher-order terms.

\begin{figure}
\epsfxsize=8.5cm
\epsffile[ 110 340 470 610 ]{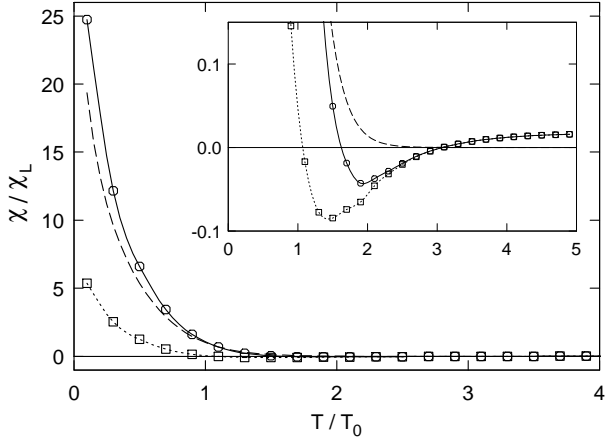}
\caption{ 
Temperature dependence of the zero-field susceptibility (solid line) for 
an ensemble of squares at $\kf a \smeq 50$. The contribution of the 
non-diagonal channel (dashed, family (11) and repetitions) exceeds that of 
the diagonal Cooper channel (dotted) at low temperatures 
(\protect{$k_{\rm B} T_0 = \hbar \vf/2\pi a$}).}
\end{figure}

{\it Diagonal Cooper channel.--} The first-order diagonal channel gives a
contribution to $\chi$ with the same dependence on $\kf
a$ and $T$ as in Eq.~(\ref{nondiag}). So, to first
order in the perturbation the difference between chaotic systems--
for which there is only the diagonal contribution-- and regular ones-- 
for which the non-diagonal contribution is also present-- is
numerical but not qualitative.

However, higher-order diagrams are essential in the diagonal
Cooper channel, as known from the theory of superconductivity
\cite{AGD,aslamazov}.  One should sum all terms 
which (i)~do not vanish upon ensemble averaging, (ii)~depend 
on $B$, and (iii)~are of leading order in $\hbar\sim
1/ \kf a$.  This yields the Cooper series shown in Fig.~1.
For instance, (iii) is checked by $\hbar$ power
counting, since a pair of Green functions scales as $ N(0)/
\hbar $, interactions as $ [N(0)]^{-1} $, and Matsubara sums
as $\hbar$.  Indeed, all terms in the series are of order $\hbar$
despite the formal expansion in $\lambda_0$.  Summing the series
yields for the diagonal contribution \cite{doubleA}
\begin{equation}
   \label{omega_diag} 
   \Omega^{(D)} = {1\over \beta} \sum_\omega {\rm Tr} \left\{ 
   \ln [1+ \lambda_0 \Sigma_{\br, \br'}^{(D)}(\omega) ] \right\} \;.  
\end{equation} 
The diagonal part $\Sigma^{(D)}$ of $\Sigma$ 
is a sum over over all trajectories longer 
than the cutoff $\Lambda_0 = \lambdaf / \pi$ [associated with the upper
bound $\Ef$ on the Matsubara sum in Eq.~(\ref{sigma})]:
\begin{eqnarray}
   \Sigma^{(D)}_{\br, \br'}(\omega)\simeq {\hbar \over
   \pi N(0) }&& \sum_{ j : {\bf r} \to {\bf r'}}^{L_j > \Lambda_0}
   |D_j |^2\, {R(2 t_j/ t_T) \over t_j} \nonumber\\
   \times\exp&&\left[ i4\pi B A_j/\phi_0\right] \times
   \exp[-\omega t_j / \hbar]\; .  
   \label{sigmaD}
\end{eqnarray} 
While we cannot diagonalize $\Sigma^{(D)}$ analytically, it has the nice
property that (except for $\Lambda_0$) all variations occur on classical
scales: rapid oscillations on the scale of $\lambdaf$ have been washed out.
The original quantum problem becomes much simpler, involving only the
``classical'' operator $\Sigma^{(D)}$; hence, we can discretize
$\Sigma^{(D)}$ with mesh size larger than $\lambdaf$ and then compute
$\Omega^{(D)}$ numerically.

We have performed this computation for the square billiard, obtaining the
dotted  curve of Fig.~3 for $\chi (T)$.  In this curve, we can distinguish
three regimes. At low-temperature $\chi^{(D)}$ is {\it paramagnetic} and
decays on a scale similar to the non-diagonal contribution (dashed curve),
but has a significantly smaller amplitude. In the intermediate range,
$\chi^{(D)}$ is small and {\it diamagnetic}. Finally, at high temperatures
$\chi^{(D)}$ is again {\it paramagnetic}, but very small. This is naturally
understood by associating each regime with an order in the perturbation
series.  The low-temperature part corresponds to the first-order term
(orbits of the type in Fig.~2, left) which is exponentially suppressed by
the temperature factor $R$ when $L_T$ becomes smaller than the shortest
closed orbit.  At this point the second-order term, due to closed paths of
two trajectories connected by interactions, takes over. There is no minimum
length of these paths, and hence the second-order term is less rapidly
suppressed by temperature. For repulsive interactions, the sign is opposite
of the first-order term, thus the sign change in $\chi^{(D)}$.  At even
higher temperatures once $L_T\ll a$, this term is a surface contribution and
the third-order term takes over. The latter is a bulk contribution
\cite{aslamazov} since with three interactions flux can be enclosed without
bouncing off the boundary.

{\it Renormalization scheme.--} This interpretation of Fig.~3 must, however,
be reconsidered. First, as noted above the first-order diagonal and
non-diagonal contributions are of similar magnitude while the
low-temperature magnitudes in Fig.~3 are different. Second, one observes
numerically that the terms in the perturbation series increase in magnitude
with order: one is not in the radius of convergence of perturbation theory
but in its analytical continuation. Both facts contradict the above
picture.

However, the interpretation is valid once the interaction entering the
diagonal contribution is replaced by a renormalized interaction. To see
this, we propose a simple renormalization scheme where
integration over short trajectories yields a decreased effective
coupling constant. To that end consider a new cutoff $\Lambda$ 
larger than $\Lambda_0$ but much smaller than any other 
characteristic length ($a$, $L_T$, or $\sqrt{\phi_0 / B}$). For  each path
$j$ joining $\br$ to $\br'$ with $L_j > \Lambda$, let 
$\Sigma^j_{\br, \br'}$ denote its contribution to 
$\Sigma^{(D)}_{\bf r, \bf r'}$ and define 
\begin{eqnarray}
   \label{tilde} 
   \tilde\Sigma^j_{{\bf r},{\bf  r'}}
   \equiv&&\Sigma^j_{{\bf r},{\bf r'}} - \lambda_0
   \int d {\bf r}_1  \ \Sigma^j_{{\bf r},{\bf r}_1}  
   \hat \Sigma_{{\bf r}_1,{\bf r'}} \nonumber\\
   &&+\lambda_0^2\int  d {\bf r}_1 \
    d {\bf r}_2 \  \Sigma^j_{{\bf  r}, {\bf r}_1} \hat  \Sigma_{{\bf  r}_1, 
   {\bf  r}_2}\hat \Sigma_{{\bf  r}_2,  {\bf r'}}+\ldots \; .  
\end{eqnarray}
where the ${\bf r}_i$ integration is over $\Lambda_0\!<\!|{\bf r}_{i-1}-
{\bf r}_i|\!
<\!\Lambda$ (with ${\bf r}_0\!\equiv\!{\bf r'}$). 
$\hat\Sigma_{{\bf r}_1,{\bf r'}}$ is defined by Eq.~(\ref{sigmaD}) but with 
the sum restricted
to ``short'' trajectories with lengths in the range $[\Lambda_0, \Lambda]$;
$\Sigma^j_{{\bf r},{\bf r}_1}$ is obtained from $\Sigma^j_{{\bf r},{\bf r'}}$
by continuously deforming trajectory $j$.
To avoid the awkward $\ln$ in Eq.~(\ref{omega_diag}), we introduce
$\Gamma \smeq (1/\beta)\sum_\omega{\rm Tr}\, 
[1+\lambda_0\Sigma^{(D)}_{\bf r, \bf r'} (\omega)]^{-1}$, 
from which $\Omega^{(D)}$ can be derived through
\begin{equation} 
   \label{integral}
   \Omega^{(D)}(\lambda_0)         =        \int_0^{\lambda_0}    \frac{d
   \lambda'_0}{\lambda'_0}\,   \Gamma(\lambda'_0)  \;  .  
\end{equation}   
Replacing $\Sigma$ by $\tilde \Sigma$ in $\Gamma$ amounts  to a reordering
of  the perturbation expansion of $\Gamma$ in which short paths are
gathered into lower-order terms. Moreover, if $L_j  \gg \Lambda$, 
small variations in the spatial arguments do not modify  noticeably the 
characteristics of  $\Sigma^j$. Approximating
$\Sigma^j_{\br,\br_1}$ by $\Sigma^j_{\br,\br'}$   in 
Eq.~(\ref{tilde}) and using 
${\hat \Sigma_{{\bf r}_1, {\bf r'}}} \simeq 1/4\pi | \br_1 - \br' |^2$
valid for short paths,
we obtain
\begin{equation}
	\lambda_0 \, \tilde \Sigma^j_{{\bf r}, {\bf r'}} \simeq
	{ \lambda_0 \Sigma^j_{{\bf r}, {\bf r'}}\over 1 +
	\lambda_0
	\int d{\bf r_1}
	\hat \Sigma_{{\bf r}_1, {\bf r'}}}
	\simeq
        \lambda(\Lambda)\,\Sigma^j_{{\bf r}, {\bf r'}}
\end{equation}
where the running coupling constant is defined by 
$\lambda(\Lambda) =\lambda_0/[1 +(\lambda_0/2)\ln(\Lambda/\Lambda_0)]$.  
Therefore, these successive steps amount to a change of both the
coupling constant and the cutoff (since now trajectories shorter than
$\Lambda$ must be excluded) without changing $\Gamma$; that is,
\begin{equation}
   \label{RGgamma}
   \Gamma (\Lambda_0, \lambda_0) = \Gamma \left(\Lambda,
   \lambda(\Lambda)\right) \; .  
\end{equation} 
Through Eq.~(\ref{integral}), this renormalization scheme can be applied 
to $\Omega^{(D)}$, and so to the average susceptibility.

In this way, we have eliminated the last ``quantum scale'' $\Lambda_0$ from
the definition of $\Sigma^{(D)}$: $\Lambda$ can be made much larger than
$\lambdaf$ while remaining smaller than all classical lengths. Furthermore,
it is qualitatively reasonable that the perturbation series of
$\Omega^{(D)}$ becomes convergent when $\Lambda$ is of order $a$, since by
this point the spread in length scales causing the divergence has been
eliminated. We have checked that this is true numerically, although this is
at the border of the range to which $\Lambda$ can be pushed with a reliable
quantitative answer.  The interpretation of Fig.~3 is therefore correct,
once the coupling constant $\lambda_0 =1$ is replaced by the renormalized
one $\lambda(a)\simeq 2/\ln(\kf a)$.  The smallness of $\lambda(a)$
explains, first, why the magnitude of the diagonal contribution is reduced
below the off-diagonal one in Fig.~3 and, second, why the diamagnetic
excursion and high-temperature tail are small.

Consequently, at low $T$ ($L_T \gtrsim$ shortest periodic orbit), the
diagonal contribution is {\it parametrically} smaller than the non-diagonal
one by a factor $1/\ln(\kf a$) because higher-order correlation terms reduce
only the diagonal contribution. {\it Therefore, regular systems, for which
there is a non-diagonal contribution, show a magnetic response
logarithmically larger than generic chaotic systems, for which only the
diagonal channel is open}. For comparison, we note that the non-interacting
contribution obtained previously \cite{vop95,urj95} is of the same order as
this interaction contribution for integrable systems but smaller for chaotic
ones.

To conclude, we have shown that a semiclassical treatment allows one to
study the high-density perturbative expansion of the interaction
contribution to the grand potential for ballistic quantum dots.  This
semiclassical approach is an efficient tool to compute quantitatively the
magnetic response.  Moreover, when combined with a renormalization scheme,
it provides an intuitive picture of various features specific to the
ballistic regime. The most striking one is that the susceptibilities of
integrable and chaotic geometries scale differently with $\kf a$ because of
the presence of families of periodic orbits in the former. Another unusual
property, caused by the different $T$ dependence of different orders in the
(renormalized) interaction, is that with increasing temperature the
interaction contribution changes sign from paramagnetic to diamagnetic and
then back to paramagnetic.

RAJ, FvO, and KR thank the ITP Santa Barbara (PHY94-07194) where part of
this research was performed.  RAJ and KR acknowledge support from the
French-German program PROCOPE.  The Division de Physique Th\'eorique is
``Unit\'e de recherche des Universit\'es Paris~11 et Paris~6 associ\'ee au
C.N.R.S.''.


\begin{references}
\bibitem{localint}
This amounts to neglecting the momentum dependence of the Thomas-Fermi
potential, which in two dimensions is
$\hat U  (p) =  N(0)^{-1}  / [1   + r_s^{-1} \sqrt{1/2}  (p /\pf)  ]$.
Since the momentum transfers along the trajectories we consider are
of order $\pf$, we may replace $\hat U (p)$ by $\hat U (0)$
for typical $r_s$.
\end{references}
\end{document}